\documentclass{article}
\usepackage{arxiv}

\usepackage{amssymb}
\usepackage{amssymb,amsmath}
\usepackage{mathtools}
\usepackage{eqnarray}
\usepackage{IEEEtrantools}
\usepackage{multirow}
\usepackage{placeins}
\usepackage{booktabs}
\usepackage{xspace}

\usepackage{float}

\usepackage{graphicx}
\usepackage{subfig}
\usepackage{bm}
\usepackage[colorlinks,bookmarksopen,bookmarksnumbered,citecolor=red,urlcolor=red]{hyperref}\usepackage{soul}
\usepackage{enumerate}
\usepackage{enumitem}
\usepackage{setspace}
\usepackage{textcomp}
\usepackage{todonotes}
\graphicspath{{Figures/}}
\usepackage{txfonts}
\usepackage{dsfont}
\newlist{steps}{enumerate}{1}
\setlist[steps, 1]{label = \textbf{Step \arabic*:},leftmargin=1.3cm}
\usepackage{cleveref}
\usepackage{caption}
\newcommand{\Hfull}{ {\mathcal{H}}(s)} 
\newcommand{\Hf}{ \mathcal{H}} 

\newcommand{\Hr}{ \widetilde{\mathcal{H}}}
\newcommand{\ssA}{ \varmathbb{A}} 
\newcommand{\ssB}{ \varmathbb{B}} 
\newcommand{\ssC}{ \varmathbb{C}} 
\newcommand{\norm}[1]{\left\lVert#1\right\rVert}
\newcommand{\ssAr}{ \widetilde{\varmathbb{A}}}
\newcommand{\ssBr}{ \widetilde{\varmathbb{B}}}
\newcommand{\ssCr}{ \widetilde{\varmathbb{C}}}

\newcommand{\relerr}{\mathcal{E}_{L_2}}
\newcommand{\weightrelerr}{\mathcal{E}_{L_2}^{w}}

\newcommand{\sk}{{\small{{SK}}}\xspace}
\newcommand{\vf}{{\small{{VF}}}\xspace}
\newcommand{\frf}{{\small{{FRF}}}\xspace}

\newcommand{\ls}{{\small{{LS}}}\xspace}

\usepackage[switch, modulo]{lineno}

\usepackage[utf8]{inputenc} 
\usepackage[T1]{fontenc}    
\usepackage{hyperref}       
\usepackage{url}            
\usepackage{booktabs}       
\usepackage{amsfonts}       
\usepackage{nicefrac}       
\usepackage{microtype}      
\usepackage{cite}

\title{Estimating Experimental Dispersion Curves from Steady-State Frequency Response Measurements }

\author{V. V. N. Sriram Malladi
\thanks{Address all correspondence to this author.}\, 
\thanks{These two authors have contributed equally.} \\
 Department of Mechanical Engineering - \\ Engineering Mechanics \\
  Michigan Technological University,  \\
  Houghton, Michigan, 49931 \\
   \texttt{smalladi@mtu.edu} \\
\And 
Mohammad I. Albakri$^\dagger$ \\ 
Department of Mechanical Engineering \\ Tennessee Technological University, \\ 
Cookeville, Tennessee, 38505 \\
\texttt{malbakri@tntech.edu } \\
\And 
Manu Krishnan \\
Department of Mechanical Engineering \\  Virginia Tech \\
Blacksburg, Virginia, 24060 \\
\texttt{manukris@vt.edu } \\
\And Serkan Gugercin \\ 
Department of Mathematics \\ 
Virginia Tech \\
Blacksburg, Virginia, 24060 \\
\texttt{gugercin@vt.edu} \\
\And Pablo A. Tarazaga  \\
Department of Mechanical Engineering \\  Virginia Tech \\
Blacksburg, Virginia, 24060 \\
\texttt{ptarazag@vt.edu} \\
}

\begin{document}
\maketitle

\begin{abstract}
Dispersion curves characterize the frequency dependence of the phase and the group velocities of propagating elastic waves. Many analytical and numerical techniques produce dispersion curves from physics-based models. However, it is often challenging to accurately model engineering structures with intricate geometric features and inhomogeneous material properties.  For such cases, this paper proposes a novel method to estimate group velocities from experimental data-driven models. Experimental frequency response functions (FRFs) are used to develop data-driven models, {which are then used to estimate dispersion curves}. The advantages of this approach over other traditionally used transient techniques stem from the need to conduct only steady-state experiments. In comparison, transient experiments often need a higher-sampling rate for wave-propagation applications and are more susceptible to noise. 
 
The vector-fitting (\vf) algorithm is adopted to develop data-driven models from experimental in-plane and out-of-plane FRFs of a one-dimensional structure. The quality of the corresponding data-driven estimates is evaluated using an analytical Timoshenko beam as a baseline. The data-driven model (using the out-of-plane FRFs) estimates the anti-symmetric ($A_0$) group velocity with a maximum error of $4\%$ over a 40~kHz frequency band. In contrast, group velocities estimated from transient experiments resulted in a maximum error of $6\%$ over the same frequency band. 
\end{abstract}

\keywords{data-driven models  \and dispersion curves \and least-squares \and vector-fitting algorithm \and longitudinal and flexural models}

\section{Introduction}
Numerous applications incorporate guided waves as essential components. Many structural health monitoring algorithms, for instance, rely on guided-waves to detect and localize structural damages ~\cite{ciampa2010new,rosalie2004variation}.  These techniques depend on accurately discerning variations in guided-wave features ~\cite{lee2011time}.  However, inhomogeneity in material properties and complex geometric features makes it challenging to determine their actual attributes. As a result, there is limited applicability of these techniques to in-homogeneous systems such as reinforced-concrete floors and non-uniform composites.

Over the years, several approaches have been developed to experimentally determine dispersion curves and capture the frequency-dependent nature of wave numbers associated with different wave modes. These approaches can be broadly categorized based on the nature of testing as time-domain transient approaches or frequency-domain steady-state approaches. Traditionally, most of the dispersion estimation relies on the time domain analysis of transient measurements. Popular techniques include time-of-flight analysis ~\cite{ziola1991source}, cross-correlation techniques ~\cite{ziola1991source}, wavelet-based methods ~\cite{jeong2000wavelet}, and model-based approaches ~\cite{hall2010model}.  Depending on the structure under test, there are advantages and disadvantages of using time-domain techniques. The advantages extend from the need for minimal spatial information. For instance, estimating the time taken by a guided waveform to travel between two know spatial locations is sufficient to establish its phase and group velocity. However, experimental noise and contamination of measured signals from reflections hamper the reliability of the transient approaches. 

Frequency-domain approaches, on the other hand, such as the spatial Fourier transform~\cite{van2018measuring}, the Bayesian estimation~\cite{souza2020bayesian}, and the in-homogeneous wave correlation \cite{berthaut2005k} rely on the steady-state dynamic response in the form of Frequency Response Functions (FRFs) and Operating Deflection Shapes (ODSs). The advantages of steady-state approaches stem from the relative ease of measuring FRFs and ODSs as opposed to propagating waveforms ~\cite{chaigne2014structural,liu2001elastic,doyle1989wave}. In time-domain tests, accurate measurement of a  propagating wave necessitates a sampling rate of at least 10-15 times the maximum frequency content of the waveform. In contrast, the need for strict temporal resolution is relaxed with steady-state dynamical measurements where a sampling rate of 2 - 2.5 times the maximum frequency of interest is sufficient. Generally, repeatability is higher with steady-state measurements, and averaged FRFs have superior signal-to-noise ratio compared to transient time-domain measurements. However, successful implementation of the aforementioned frequency-domain approaches requires fine spatial resolution to avoid spatial aliasing. As a result, the steady-state response needs to be measured at a large number of points, which adds to the cost and complexity of the underlying experiments.

In a recent effort by the authors~\cite{albakri2020estimating}, a novel data-driven-based approach that brings together the advantages of both the time-domain and frequency-domain testing has been developed. This approach uses FRFs to develop a data-driven dynamical model for the structure under test. Dispersion curves are then determined using the developed model by applying transient time-domain techniques to numerically simulated waveforms. Thus, the advantages of steady-state experimentation is combined with the relaxed spatial resolution requirements of the transient approaches. While there is a plethora of approaches for constructing data-driven rational approximations from measured data (see, e.g., \cite{aca90,ALI18,mayo2007fsg,karachalios2020loewner,AntBG20} for interpolatory methods;~\cite{hokanson2017projected,gustavsen1999rational,Drmac-Gugercin-Beattie:VF-2014-SISC,berljafa2017rkfit,hokanson2018least} for least-squares based approaches; and~\cite{nakatsukasa2018aaa,gosea2020algorithms,rodriguez2020p,nakatsukasa2020algorithm} for rational approximations that combine the interpolatory and least-squares frameworks), the least-squares based vector-fitting (\vf) method \cite{gustavsen1999rational} was the method of choice in \cite{albakri2020estimating} and the ability of the FRF-based data-driven models to simulate the transient dynamics was established. Through numerical experiments, the capabilities of this new approach has been demonstrated where accurate estimates of dispersion curves corresponding to longitudinal and flexural wave modes in one-dimensional structures were obtained.     

This paper extends this investigation by tailoring the \vf~algorithm to experimentally measured data. Other rational data-driven modeling frameworks such as those listed above can be employed here as well; {however this work uses the \vf~algorithm to experimentally validate and test the authors' earlier work \cite{albakri2020estimating}, which also employed \vf.} Unlike numerically generated counterparts, experimentally measured FRFs present several challenges to data-driven modeling. This includes noise, multi-mode excitation (resulting in mixed axial, flexural, and torsional FRFs), and non-ideal boundary conditions. The current work addresses practical issues in fitting such noisy data and provides a \vf framework to build data-driven models from experimentally measured FRFs. The impact of such artifacts on data-driven models, their ability to capture the dynamic system's response, and resulting predictions of dispersion curves is investigated. The current paper is structured as follows: Section \ref{Sec:Setup} describes the experimental setup, and the test procedure followed to generate in-plane and out-of-plane FRF. A brief overview of the nonlinear least-squares problem and the iterative vector fitting techniques is presented in Section \ref{Sec:vf}.  Implementation of the VF technique to develop out-of-plane and in-plane data-driven models is discussed in Section \ref{Sec:datadriven}. The performance of these models in estimating dispersion curves is the topic of discussion in Section \ref{Sec:dispersion}. Towards the end, an overview of the results is summarized along with final remarks for future work.


\section{Experimental Setup and Test Procedure}
\label{Sec:Setup}
\begin{figure}[!htb]
\centering
\includegraphics[trim=0cm 3cm 0cm 3.5cm, width=\textwidth]{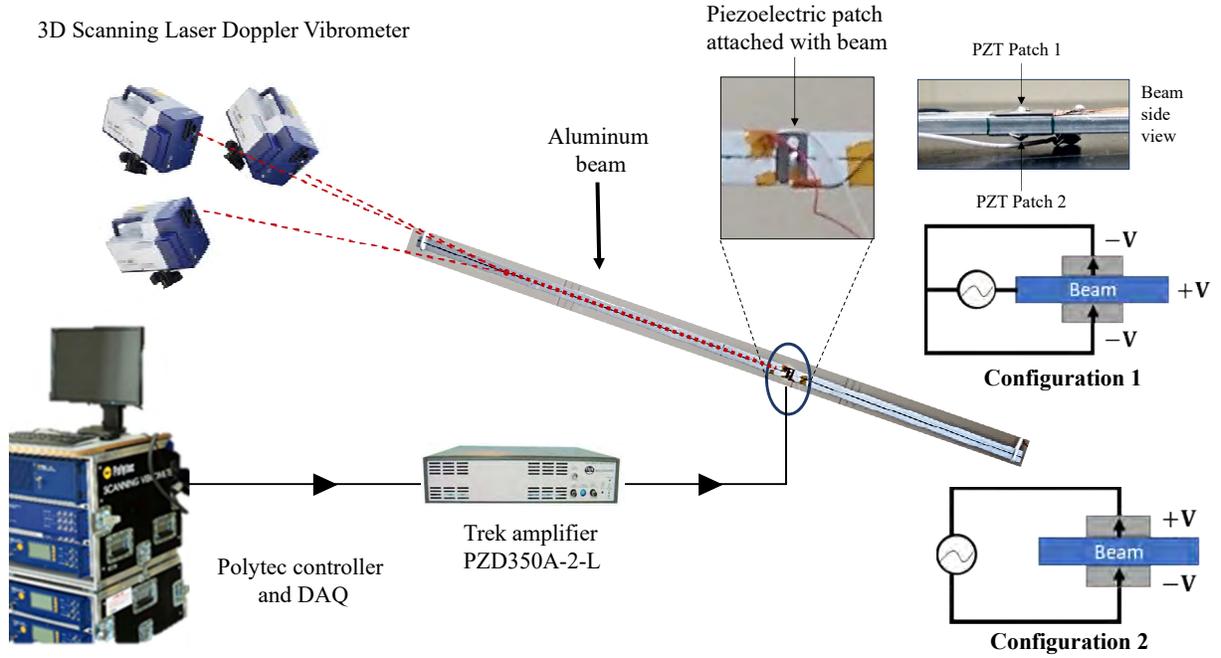}
\caption{\label{fig:exptstp} Experimental setup measuring the in-plane and the out-of-plane vibrations of the aluminum beam when excited with two piezoceramics}
\end{figure}

The experimental set up used to measure in-plane, and out-of-plane FRFs~ is depicted in Figure \ref{fig:exptstp}. A $48~in.$ long aluminum beam with a rectangular cross-section of $ (1~in. \times 1/8~in.$) has been selected for this study. Free-free boundary conditions are approximated by suspending the beam under test with two bungee cords. Two piezoceramic wafers (PZT-5H) of dimensions $L_p\times W_p$ $ (1~in. \times 1/8~in.$) are bonded to the upper and lower faces of the beam, $18.5~in.$ from its left end. The poling direction of the piezoceramics (PZTs), along with the two possible electrical configurations, are shown in the figure. Configuration 1 is designed to amplify the net bending moments generated by actuating the PZTs while cancelling in-plane forces. This allows for the measuring out-of-plane FRFs~with minimal contribution from in-plane deformations. Configuration 2, on the other hand, is designed to neutralize bending moments and amplify in-plane forces to facilitate the measurement of in-plane FRFs. However, perfect decoupling between in-plane and out-of-plane deformations is not possible due to PZTs mounting imperfections and misalignment. Thus, a combination of vibrational responses corresponding to longitudinal, flexural, and torsional modes always appear in experimental measurements. An in-depth experimental modal analysis would differentiate these modes of vibration. The vector-fitting algorithm can conveniently include the dynamics corresponding to all of these vibrational modes in the data-driven models. 

A 3D Scanning Laser Doppler Vibrometer (3D-SLDV) is used to measure the beam's dynamic response when excited with PZTs under both configurations. The 3D-SLDV has three laser vibrometers that simultaneously measure the velocity response at every scanning point on the beam. The Polytec software then analyzes the measurements and transforms the data to compute the velocity components along the three perpendicular axes. As a result, it is possible to measure both in-plane and out-of-plane responses of the beam. 3D-SLDV also triangulates the geometric coordinates of the points as it measures the response along the beam. This provides accurate information about the coordinates and distance between measurement points, which is later used for estimating dispersion curves. It is important to note that there is an uncertainty of $0.5~mm$ while measuring the coordinates of the measurement points, which will be reflected in group velocity calculations. In the present set of experiments, the velocity measurements are recorded at 19 equally spaced scan points along a straight line of about $18~in$, starting $2~in$ away from the PZTs.

Using the experimental setup shown in Figure \ref{fig:exptstp}, two sets of experiments have been conducted. The first set targeted the steady-state dynamic response of the beam in the form of FRFs. For this purpose, the beam is excited with a swept-sine signal covering the frequency range of $0-50 kHz$. These FRFs are then used to construct the data-driven model of the beam. A sampling frequency of $128~kS/s$ is used for this experiment, and the measured FRFs have a frequency resolution of $0.25~Hz$.  At each point, the mean FRF and the mean coherence estimates are determined for a set of 10 repeated trials. The second set of experiments targeted the transient dynamic response of the beam. For this purpose, the beam is excited with a $2-$ or $4-$cycle, amplitude-modulated, tone burst with a center frequency ranging from $1~kHz$ to $50~kHz$. This set of experiments aims to provide the data necessary to directly measure the beam's dispersion curves, which is used for comparison purposes. A sampling rate of $1.28~MS/s$ is used for this set of experiments, which is ten times the rate used for measuring FRFs~in the first set. For both sets of experiments, in-plane and out-of-plane responses are excited and measured using the configurations mentioned earlier. Excitation signals are generated using the Polytec function generator and a Trek (PZD350A-2-L) power amplifier, as shown in the figure.


\section{Data-driven frequency response (rational) approximation based on the vector-fitting algorithm}
\label{Sec:vf}
During the experiments, the Polytec data acquisition system provides the frequency response of the beam  at $n_{out} = 19$ locations ($v(t) = [v_1(t),...,v_{n_{out}}(t)]^T$) to a voltage input $f(t)$ applied to the PZT. 
Let $V(s)$ and $F(s)$ denote the Laplace transforms of 
$v(t)$ and $f(t)$, respectively. Then, the corresponding frequency response function $\Hfull$ satisfies $V(s) = \Hfull F(s)$.
Our experimental set-up samples this frequency response function $\Hfull$ at multiple frequencies $s_j=\dot{\imath}\omega_j$ for $j=1,2,\ldots,N$. In our experiments, $N=204800$. 
The sampled estimate of a single-input/single output (SISO) FRF~at a frequency $s=\dot{\imath} \omega$ is then calculated using  $\mathcal{H}_1$ estimator and/or $\mathcal{H}_2$ estimator \cite{ewins1984modal} given by, 
\begin{eqnarray}
\mathcal{H}_1(\dot{\imath} \omega) &= \dfrac{\mathcal{G}_{vf}(\dot{\imath} \omega)}{\mathcal{G}_{ff}(\dot{\imath} \omega)}  
\qquad\mbox{and}\qquad\mathcal{H}_2(\dot{\imath} \omega) &= \dfrac{\mathcal{G}_{vv}(\dot{\imath} \omega)}{\mathcal{G}_{fv}(\dot{\imath} \omega)}  ,
\end{eqnarray}
where the cross-power spectral densities and auto-power spectral densities are defined as $\mathcal{G}_{vf}(\dot{\imath} \omega) = E~\big[V(\dot{\imath} \omega)F(\dot{\imath} \omega)^*\big]$, $\mathcal{G}_{vv}(\dot{\imath} \omega) = E~\big[V(\dot{\imath} \omega)V(\dot{\imath} \omega)^*\big]$ and $\mathcal{G}_{ff}(\dot{\imath} \omega) = E~\big[F(\dot{\imath} \omega)F(\dot{\imath} \omega)^*\big]$, respectively; and $E~\big[~.~\big]$ denotes the estimator function and $\{\cdot\}^*$ is the complex-conjugate of a the matrix.  In the absence of noise, $\mathcal{H}_1(\dot{\imath} \omega)$and 
$\mathcal{H}_2(\dot{\imath} \omega)$  
result in  identical FRFs. However, in the presence of noise in either the input or the output measurements, both methods approximate the inherent dynamics of the structure. Generally, the $\mathcal{H}_1(\dot{\imath} \omega)$ formulation is sensitive to noise on the input signal $f(t)$ and underestimates $\mathcal{H}(\dot{\imath} \omega)$. On the contrary,  the $H_2(\dot{\imath} \omega)$ formulation is sensitive to noise on the output signal $v(t)$ and overestimate the true FRF. Therefore, the quality of the measured FRFs can be determined by the coherence function $\gamma^2(\dot{\imath} \omega)$ defined as the ratio of the two formulations, 
\begin{align}
    \gamma^2{(\dot{\imath} \omega)} = \dfrac{\mathcal{H}_1(\dot{\imath} \omega)}{\mathcal{H}_2(\dot{\imath} \omega)} = \dfrac{|\mathcal{G}_{vf}(\dot{\imath} \omega)|^2}{\mathcal{G}_{vv}(\dot{\imath} \omega)\mathcal{G}_{ff}(\dot{\imath} \omega)}.
\end{align} 
Typically, coherence values lie in the range of $[0~-~1]$, where coherence value close to unity represents a high signal-to-noise ratio. See \cite{ewins1984modal} for further details. For multi-input/multi-output (MIMO) systems as in the present case, the FRF formulation can be written as
\begin{align}
\mathcal{H}(\dot{\imath} \omega) &= \mathcal{G}_{vf}(\dot{\imath} \omega)\mathcal{G}_{ff}^{-1}(\dot{\imath} \omega).
\end{align} 
The output cross-spectral density can be determined using two methods
\begin{align}
    \mathcal{G}_{vv}(\dot{\imath} \omega)=\mathcal{H}(\dot{\imath} \omega)\mathcal{G}_{ff}(\dot{\imath} \omega)\mathcal{H}(\dot{\imath} \omega)^{*}\quad\text{and}\quad
    \tilde{\mathcal{G}}_{vv}(\dot{\imath} \omega)=\mathcal{H}(\dot{\imath} \omega)\mathcal{\mathcal{G}}_{fv}(\dot{\imath} \omega).
\end{align}
Although these two formulation are  equivalent
theoretically,
experimental noise results in numerical discrepancies.  The coherence function of a MIMO system is defined as the correlation coefficient between each output ($v_i(t)$) to the input ($f(t)$). Therefore, the multiple coherence function $\gamma^2_{v_i:f}(\dot{\imath} \omega)$ is the ratio of the diagonal elements of $\tilde{\mathcal{G}}_{vv} $ and $\mathcal{G}_{vv}$, i.e.,
\begin{align}
    \gamma^2_{v_i:f}(\dot{\imath} \omega)&=\dfrac{\tilde{\mathcal{G}}_{v_iv_i}(\dot{\imath} \omega)}{{\mathcal{G}}_{v_iv_i}(\dot{\imath} \omega)}.
\end{align}
This work utilizes the $\mathcal{H}_1$ estimates of the FRFs~to develop Single-Input Multi-Output (SIMO) data-driven models. For simplicity,
the subscript is dropped and the  experimental FRFs~ are denoted by $\Hf( \dot{\imath} \omega)$ throughout the paper. 
Therefore, the experimental \frf~data are
\begin{IEEEeqnarray}{CC} 
\Hf( \dot{\imath} \omega_j)\in \mathbb{C}^{19\times1}~~\mbox{for}
~~j=1,2,\ldots,204800. \label{frfdata}
\end{IEEEeqnarray}
Given the \frf~data in \eqref{frfdata}, the objective is to
construct a low-order rational approximation (a reduced-order frequency response function) $\Hr(s)$ that approximates the data, and thus the true frequency response function $\Hfull$, in an appropriate measure.  This measure will be clarified below.
Once $\Hr(s)$ is obtained, it could be used to study the underlying dynamics and determine the system parameters such as the poles and the residues of the system. Generally, most of the modal-identification
techniques in the frequency domain start by assuming a rational function to fit the FRFs.

A rational function  could be represented in many equivalent forms. To simplify the presentation, for time being,
assume that $\Hfull$ is a SISO system; thus so is the synthesized \frf, $\Hr(s)$.
Let $\Hr(s)$ have order $r$. Then, 
it could be simply written as 
\begin{IEEEeqnarray}{CC}
\Hr(s)= \frac{n(s)}{d(s)} =
\frac{\alpha_0 + \alpha_1 s + \cdots+ 
\alpha_{r-1} s^{r-1}}{\beta_0 + \beta_1 s + \cdots+ 
\beta_{r-1} s^{r-1}+ s^r},\label{eq:rationalform}
\end{IEEEeqnarray} 
where the numerator $n(s)$ and denominator $d(s)$
are polynomial of  degree  $r-1$ and $r$, respectively,.
The numerator coefficients $\{\alpha_i\}$ and  
the denominator coefficients $\{\beta_i\}$ fully describe it. Since, in our structural dynamics, the underlying frequency response function is strictly proper, i.e., the order of the denominator is strictly greater than that of the numerator, $\Hr(s)$ in \eqref{eq:rationalform} is chosen to have the same form. This is not a restriction and the general case can be easily handled, but is not relevant to our set-up.

The unknowns coefficients $\{\alpha_k\}$ and $\{\beta_k\}$ are  typically estimated through the least-squares (\ls) approach. In other words, given the \frf~data in \eqref{frfdata}, the goal is to construct 
$\Hr(s)$ such the \ls error  
\begin{IEEEeqnarray}{CC}
J = \sum_{j=1}^N w_j\left| \Hf(\dot{\imath} \omega_j) - \Hr(\dot{\imath} \omega_j) \right|^2\label{eq:lsfit}
\end{IEEEeqnarray} 
is minimized where $w_j$'s are the weights corresponding to the frequency $\omega_j$.  This \ls problem is nonlinear due to its nonlinear dependence on the denominator coefficients. The first step in most techniques popular in the modal-testing community ~\cite{allemang1998unified}
is to linearize this nonlinear problem via
$$
 \Hfull - \Hr(s)  =  \Hfull - \frac{n(s)}{d(s)}
~~\rightsquigarrow~~  d(s)\Hfull - n(s).  
$$
Then, instead of the original nonlinear \ls error 
in \eqref{eq:lsfit}, one minimizes, as done in, e.g., \cite{formenti2002parameter}, the linearized \ls error
\begin{IEEEeqnarray}{CC}
J_{lin} = \sum_{j=1}^N w_j \,\big| d(\dot{\imath} \omega_j) \Hf(\dot{\imath} \omega_j) - n(\dot{\imath} \omega_j)  \big|^2. \label{eq:linlsfit}
\end{IEEEeqnarray} 
Similarly, other modal identification techniques linearize the rational-polynomial to apply a \ls procedure to estimate coefficients \cite{maia1997theoretical,ewins1984modal, guillaume2003poly}. Allenmang et.al.~\cite{allemang1998unified,allemang2006complete} have developed a unified matrix polynomial framework to identify the similarities and differences between multiple modal identification techniques available in the literature. 
These approaches convert the nonlinear least-square problem
\eqref{eq:lsfit} 
to a linear one \eqref{eq:linlsfit} and the solution is obtained, without any iteration, in one step by solving the resulting linear problem.
In this work, to better handle the non-linearity, an iterative procedure will be employed: The nonlinear least-problem will be converted to solving a \emph{sequence} of linear least-square problems.

Using $\Hr(s) = n(s)/d(s)$,
the cost function $J$ in  \eqref{eq:lsfit}
 can be written as  \begin{equation} 
     J = \sum_{j=1}^N w_j\left| \Hf(\dot{\imath} \omega_j) - \Hr(\dot{\imath} \omega_j) \right|^2
     = \sum\limits_{j=1}^N w_j\left| \dfrac{\Hf(\dot{\imath} \omega_j) d(\dot{\imath} \omega_j) - n(\dot{\imath} \omega_j)}{d(\dot{\imath} \omega_j)} \right|^2. \label{NLS}
  \end{equation}
The linearization approach mentioned above amounts to ignoring the denominator in this expression to reach a linear \ls problem. 

Instead, Sanathanan and Koerner \cite{Sanathanan-Koerner-1963} proposed  a technique  in which the denominator is incorporated into the problem, leading to an iterative algorithm solving a sequence of linearized \ls problem.  In this approach, as opposed to ignoring the denominator in \eqref{NLS}, one replaces it with the denominator from the previous step. 

Let $\Hr^{(k)}(s) = n^{(k)}(s)/d^{(k)}(s)$ denote the synthesised/identified model at the $k^\text{th}$ step of the Sanathanan and Koerner (\sk) iteration. Starting with the initial guess $d^{(0)}(s) \equiv 1$, the \sk iteration, at the $k^\text{th}$ step, minimizes the linear \ls error 
\begin{equation}\label{eq:relaxedNLS}
\displaystyle \sum_{j=1}^N w_j~ \left| \frac{n^{(k+1)}(\dot{\imath} \omega_j) - d^{(k+1)}(\dot{\imath}\omega_j)\Hf(\omega_j)}{d^{(k)}(\jmath\omega_j)} \right|^2
\end{equation}
to solve for $n^{(k+1)}(s)$ and $d^{(k+1)}(s)$ and then set $\Hr^{(k+1)}(s) = n^{(k+1)}(s)/d^{(k+1)}(s)$. Note that 
 the \ls~problem in \eqref{eq:relaxedNLS} is linear in the variables $n^{(k+1)}(s)$ and $d^{(k+1)}(s)$. The iteration continues  until a desired convergence tolerance is achieved. 
 Readers are referred to \cite{vayssettes2012iterative,vayssettes2015new}, where 
 the \sk iteration in conjunction with a Gauss-Newton based optimization approach \cite{nocedal2006numerical} is employed to fit MIMO FRFs in the modal-testing community.

Writing $\Hr(s)$ as a ratio of two polynomials is only one way of representing a rational function. A mathematically equivalent but numerically more stable representation is  the so-called barycentric form \cite{berrut06barycentric} where $\Hr(s)$ is represented as a ratio of two rational functions. The Vector-Fitting (\vf) algorithm
\cite{gustavsen1999rational} precisely does this; it performs the \sk iteration using the barycentric form. As in the case of the \sk iteration, \vf is also an iterative procedure.
At the $k^\text{th}$ step, $\Hr(s)$ is parameterized as
\begin{equation}\label{eq:H_r(k)}
\Hr^{(k)}(s) =
\frac{\tilde{n}^{(k)}(s)}{\tilde{d}^{(k)}(s)} =
\frac{\sum_{i=1}^r \phi_i^{(k)}/(s-\lambda_i^{(k)})}{1 + \sum_{i=1}^r \psi_i^{(k)}/(s-\lambda_i^{(k)})},
\end{equation}
where $\phi_i^{(k)},\psi_i^{(k)}, \lambda_i^{(k)} \in\mathbb{C}$ and $\{\lambda_i^{(k)}\}$ are an arbitrary set of mutually distinct points. 
Therefore, in \vf, the model $\Hr(s)$ is parameterized by
$\phi_j^{(k)}$, $\psi_j^{(k)}$, and $\lambda_j^{(k)}$. The \sk iteration, in this new barycentric form, can be considered as running the iteration \eqref{eq:relaxedNLS} for \emph{fixed} $\{\lambda_i^{(k)}\}$.  \vf takes advantage of the fact that $\{\lambda_i^{(k)}\}$ are free and can be updated in every step of the iteration. 
It does so by choosing the zeros of the numerator  $ \tilde{d}^{(k)}(s) = 1 + \sum_{j=1}^r \psi_j^{(k)}/(s-\lambda_j^{(k)})$ in \eqref{eq:H_r(k)} as the next set of 
$\{\lambda_i^{(k)}\}$.  Similar to the \sk iteration, in every step of \vf, a linear \ls problem is solved to update the parameters $\phi_i^{(k)},\psi_i^{(k)}$, and $\lambda_i^{(k)} \in\mathbb{C}$. The iteration and updates continue until convergence, upon which, due to the updating scheme,  $\tilde{d}^{(k)}(s) \to 1$, yielding the final form of the approximant
\begin{IEEEeqnarray}{lCr}
   \Hr(s) = \sum_{k=1}^r \frac{\phi_k}{s-\lambda_k},\label{eq:rf}
\end{IEEEeqnarray}
in the pole-residue form: the sum of `$r$' modal components, i.e., the poles ($\{\lambda_i\}$) and residues ($\{\phi_i\}$).  Therefore, even though  $\lambda_i^{(k)}$'s are not the poles of $\Hr(s)$ in \eqref{eq:H_r(k)} initially, upon convergence, they do become the poles of $\Hr(s)$. Due to this connection, we will call $\{\lambda_i^{(k)}\}$ as the poles; but this should be understood so in the context of convergence.  
A brief remark on the convergence of \vf is in order:  Even though one can construct examples where \vf does not converge \cite{lefteriu2013convergence}, 
when $\{\lambda_i\}$ are initialized appropriately, it usually converges
quickly. Proper initialization becomes even more important if the order $r$ grows 
modestly. This is indeed the case in our setting and will be discussed in more detail below. For further details on \vf, see \cite{gustavsen1999rational,gustavsen2006improving,Chinea-G.Talocia-2011,Drmac-Gugercin-Beattie:VF-2014-SISC,drmac2015vector,grivet2015passive} and the references therein.
   In the modal-testing community, a recent technique, called Maximum Likelihood estimation of a Modal Model ~\cite{el2019mlmm,el2016incorporating},
 uses the pole-residue parametrization 
 of the transfer function as in~\eqref{eq:rf}, i.e., takes the residues $\phi_k$ and poles $\lambda_k$ as the variables, and 
 employs the Levenberg-Marquardt~\cite{more1978levenberg} scheme to fit these parameters to \frf samples. Either using the pole-residue or the barycentric  form, one can indeed employ a variety of well-established optimization techniques for the resulting nonlinear LS problems; see for example \cite{golub1973differentiation,hokansonphd,hokanson2017projected} and the references therein.

The formulation above assumed, for simplicity, a SISO model. In our experimental set-up, we have a SIMO system with $n_{out}=19$ outputs. The theoretical analysis above directly extends to our SIMO setting. For the 
data in  \eqref{frfdata}, the \ls error in \eqref{eq:lsfit} is now given by
\begin{IEEEeqnarray}{CC}
J = \sum_{j=1}^N w_j\left\| \Hf(\dot{\imath} \omega_j) - \Hr(\dot{\imath} \omega_j) \right\|_2^2,\label{eq:lsfitsimo}
\end{IEEEeqnarray} 
where $\| \cdot \|_2$ denotes the $2$-norm. Then, the only revision in the barycentric form in \eqref{eq:H_r(k)} is that $\phi_i$ is now a column vector of size $19\times 1$. Therefore, upon the convergence of \vf, we have
\begin{equation} \label{Hr_in_pr}
\Hr(s) = \sum_{j=1}^r \frac{\phi_j}{s-\lambda_j}~~\mbox{where}~~\phi_i \in \mathbb{C}^{19\times 1}
~~\mbox{and}~~\lambda_i \in \mathbb{C} ~\mbox{for}~i=1,\ldots,r.\end{equation}
It is usually preferred to have a state-space formulation for frequency-response function. Therefore, we will represent  
\eqref{Hr_in_pr} equivalently in a real state-space form, i.e,,
\begin{equation} \label{Hr_in_ss}
\Hr(s) = \sum_{j=1}^r \frac{\phi_j}{s-\lambda_j} =
\ssCr(s I - \ssAr)^{-1}\ssBr
~~\mbox{where}~~\ssAr \in \mathbb{R}^{r\times r},~~
\ssCr \in \mathbb{R}^{19\times r},~~\mbox{and}~~\ssBr \in \mathbb{R}^{r}.
\end{equation}
Note that the realness of the state-space realization in \eqref{Hr_in_ss} is guaranteed by implicitly including the conjugate data, i.e., $\{\Hf(-\dot{\imath} \omega_j)\}$, in the \frf data.

\section{Discussion on Fitting Frequency Response Functions}
\label{Sec:datadriven}
This section investigates the performance of two SIMO data-driven models. While one simulates the flexural response, the other simulates the beam's in-plane response over a frequency band of $0-50~kHz$.  There is a significant number of natural frequencies in the frequency range of interest. Notably, there are more out-of-plane natural frequencies for the current 1D-structure. Consequently, it is challenging to work with out-of-plane FRFs. A multi-step modeling procedure is employed to fit the experimental out-of-plane data-set. On the other hand, it is feasible to vector fit all in-plane FRFs in a single setting.

\subsection{Implementation procedure of VF algorithm for fitting FRFs over a large frequency bandwidth}

In the current work, a data-driven model is developed over a frequency bandwidth of [$0-50~kHz]$. Typically, natural frequencies of dispersive structures are unevenly distributed over such a broad frequency bandwidth; thus, modal density (number of poles in a frequency interval) is higher at lower frequencies than at higher frequency bandwidths. Modal density also plays a role in fitting FRFs, leading to over/underfitting parts of the frequency bandwidth as the experimental FRFs~are sampled uniformly.  Vector fitting a broad bandwidth of $50kHz$ requires many poles, resulting in a large model order $r$. These factors make it challenging to uniformly fit experimental measurements with 208400 samples for each of the 19 FRFs. Consequently, while fitting such big-datasets, the choice of initial/starting poles significantly impacts the convergence of the \vf algorithm and the quality of the final rational approximations. Therefore, a modified approach is considered for pooling a rich choice of starting poles.

First, the total frequency bandwidth is divided into multiple bands, and the \vf algorithm is run independently for each sub-band. This approach has several advantages. As the size of the FRF data-set on individual sub-bands is drastically smaller, a low-order model can appropriately fit the FRFs, which, in turn, significantly simplifies the pole-initialization step for each band.  The model order (the number of poles in each band) is determined (approximated) based on the number of peaks/phase-change in each sub-band. Additionally, different weighting functions can be applied to fit resonant peaks and anti-resonant valleys uniformly.  Once each sub-bands are  fitted via \vf, all the fitted poles are concatenated to form a larger-set of poles. Care is taken to avoid duplication of poles in the concatenated set. This final set of combined poles is used to initialize the \vf~algorithm to fit the entire frequency bandwidth. Due to this improved pole-initialization step, the \vf algorithm converges faster and yields high-fidelity rational approximants for the full-data set, i.e., for the full frequency band.

The quality of the fitted model is determined by comparing synthesized FRFs~with experimental FRFs. In the present work, two types of relative errors are defined for evaluating the quality of the fit, namely 
\begin{IEEEeqnarray}{lCrClCr}
   \relerr &=& 
\frac{1}{N}\sqrt{
{ \dfrac{{\displaystyle \sum_{j=1}^N  \norm{\Hf(\dot{\imath} \omega_j) - \Hr(\dot{\imath} \omega_j)}_2^2}}{\displaystyle \sum_{j=1}^N \norm{\Hf(\dot{\imath} \omega_j)}_2^2} }}&~~~~\mbox{and}~~~~&\weightrelerr &=& 
\frac{1}{N}\sqrt{
{ \dfrac{{\displaystyle \sum_{j=1}^N   w_j\norm{\Hf(\dot{\imath} \omega_j) - \Hr(\dot{\imath} \omega_j)
}_2^2}}{\displaystyle \sum_{j=1}^N  w_j\norm{\Hf(\dot{\imath} \omega_j)}_2^2} }}.\label{eq:error}
\end{IEEEeqnarray}

While $\relerr$ is the relative error between experimental \frf~-$\Hf(\dot{\imath} \omega_j)$ and the synthesized \frf~-$\Hr(\dot{\imath} \omega_j)$, the error metric $\weightrelerr$ also considers the role of the weighting  scheme ($w_j$) in fitting the FRFs.

\subsection{Out-of-plane FRFs: Vector-fitting sub-bands }

The initial focus is to fit the out-of-plane FRFs over shorter frequency bands and then move to the full frequency band. In the present case, eight sub-bands are identified such that the number of peaks in each sub-band does not exceed 30. \Cref{tab:tab_section} lists the details of these sub-bands.  This approach facilitates the handling of large data sets.  Fitting individual sub-bands yields a richer distribution of starting poles, enabling faster convergence of the VF algorithm in fitting the complete FRF. 

\begin{figure}[!htb]
\centering
\includegraphics{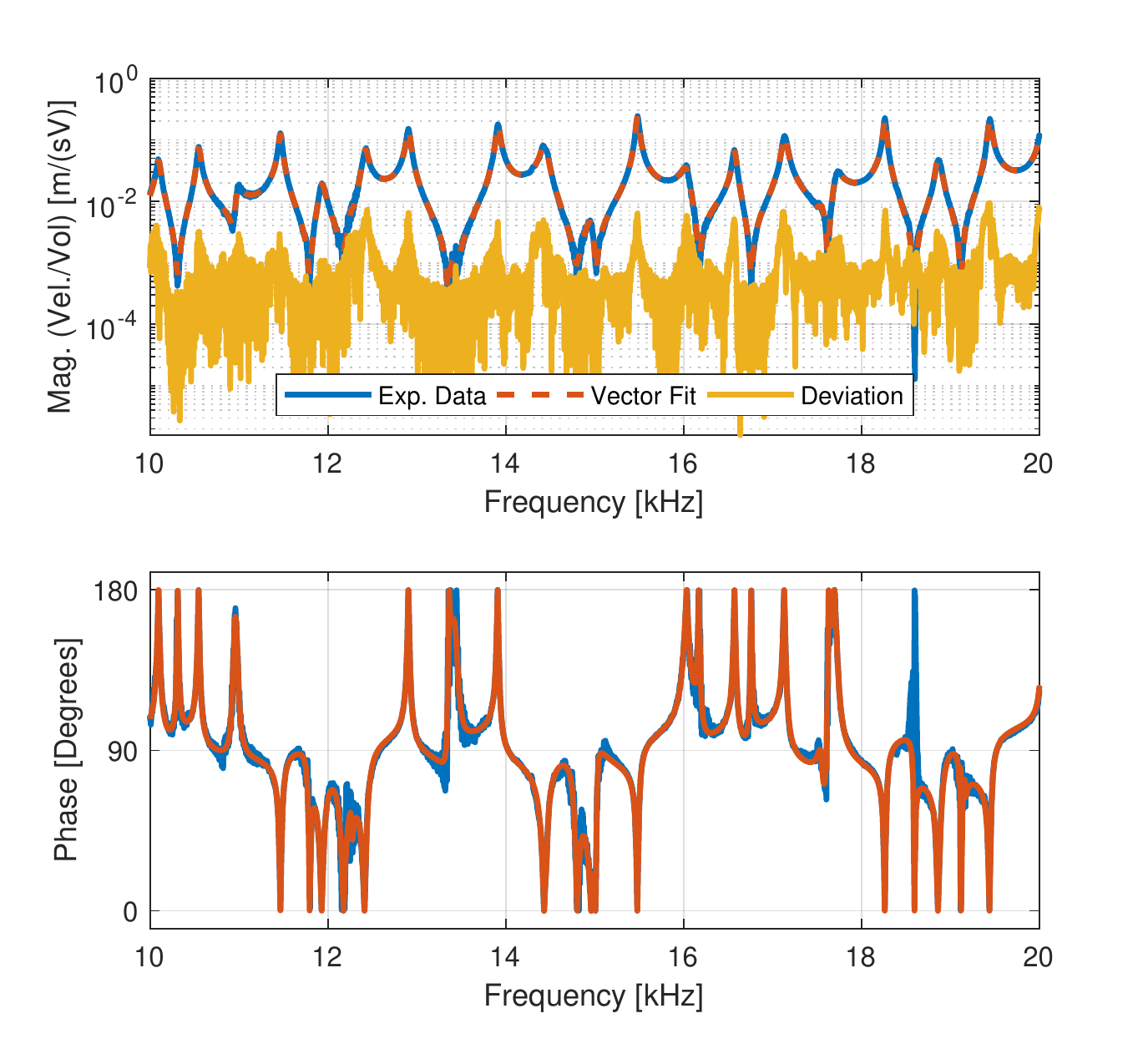}
\caption{ Comparison of the magnitude and phase plots of the experimental and the estimated FRF (at point 13) over the frequency sub-band of $10-20~kHz$. The VF estimate of the FRF is generated for the model using unit weights. }
\label{fig:Figure_vectorfit_20kHz}
\end{figure}

Peak-frequencies in each sub-band are used to initialize the VF algorithm and a unit weighting function is used to obtain an unbiased estimation of poles in each sub-band. The result is eight data-driven models that span the entire frequency band. \Cref{tab:tab_section}  summarizes the results of fitting FRFs over individual frequency sub-bands including model order and individual sub-band fitting errors ($\relerr$). \Cref{fig:Figure_vectorfit_20kHz}
compares the magnitude and the phase of the experimental FRF with the fitted FRF. The small deviation between the two FRFs highlights the quality of the fit across the frequency range.

\begin{table}[!hbt]
\centering
\caption{Partition of the complete frequency bandwidth into multiple sections. A unit weighting function is used to determine poles of sub-bands, therefore $\weightrelerr = \relerr$.}
\label{tab:tab_section}
\begin{tabular}{ @{}c|c|c|cc@{} } 
 \toprule
   Group & Frequency range  & Number of poles & $\relerr$ error  \\ 
  \hline
   1 & $ 0$ - $1kHz$      & 32 & $8.82\times 10^{-7}$ \\ 
   2 & $ 1$ - $5kHz$      & 58 & $1.82\times 10^{-7}$ \\ 
   3 & $5$ - $10kHz$      & 44 & $1.07\times 10^{-7}$ \\ 
   4 & $10$ - $20kHz$     & 56 & $7.40\times 10^{-8}$ \\ 
   5 & $20$ - $30kHz$     & 54 & $1.13\times 10^{-7}$ \\ 
   6 & $30$ - $40kHz$     & 48 & $5.81\times 10^{-8}$ \\ 
   7 & $40$ - $45kHz$     & 40 & $9.15\times 10^{-8}$ \\
   8 & $45$ - $50kHz$     & 62 & $1.08\times 10^{-7}$ \\
   \bottomrule
   \end{tabular}
\end{table}
 An ideal structural dynamics test would produce well-spaced flexural peaks in the FRFs. However, in practice, it is challenging to isolate flexural vibrations over a broad frequency band. Often, undesired artifacts show up in the FRF measurements.  While experimental irregularities induce some artifacts, the others are due to the limits of the experiment. For instance, it is interesting to examine the number of poles in each sub-band. The three frequency sub-bands within the $10-40~kHz$ frequency band require  $56$, $54$, and $40$ poles, which broadly follows the analytical perspective of a 1D-beam, i.e., the frequency spacing between consecutive natural frequencies increases. However, the  $40-50~kHz$ sub-band significantly deviates from this trend, as it needs 102 poles. 1D flexural systems do not display such sudden change in characteristics. Therefore, it is valid to assume that the structure deviates from behaving as a 1-dimensional beam above $40~kHz$, which is an important observation. In later sections, the dispersion curves further validate and establish this observation.
\subsection{Out-of-plane FRFs: Vector-fitting full-band}

 Iteratively solving the nonlinear LS problem with large data-sets is computationally challenging. Therefore, starting poles that initialize the VF algorithm are picked from the pool of poles fitting FRFs in each sub-band. The selected subset of 368 poles spans the entire $50~kHz$ band. These complex poles are chosen to correspond to the peak frequencies and other artifacts observed in the experimental FRFs. This approach avoids redundancy of starting poles, thereby avoiding over-fitting. As a result, \vf algorithm yields state-space matrices of dimensions $\ssA \in  {\rm I\!R}^{368 \times 368}$, $\ssB \in  {\rm I\!R}^{368 \times 1}$, and $\ssC \in  {\rm I\!R}^{19 \times 368}$.  

Ordinary least-squares problem formulation reduces the $L_2$ error, i.e., squared errors at all data points. Such a formulation inadvertently focuses on accurately capturing high-valued data-points. As a result, the relative error in estimating numerically smaller data-points is considerably higher. The VF algorithm also suffers from similar weaknesses and, therefore, better captures FRF characteristics at resonant frequencies than at anti-resonant frequencies. A weighted least-squares problem is one way to address this limitation. Besides, the weighting function's choice plays a vital role in estimating both resonances (peaks) and anti-resonances (valleys) of the FRFs. While accurate pole estimation correlates with peak characteristics, fitting valleys correspond with good residue (or zeros) estimation. {In this work, four weighting choices are considered. Unit weight ($w_j = 1$) and weak inverse FRF ($w_j(\dot{\imath} \omega) = 1/\sqrt{|H(\dot{\imath} \omega_j)|}$) are two of the weighing options, which are also traditionally used in VF. Having a weak inverse FRF as the weighting function enhances the data-driven model's ability to capture numerically smaller values near anti-resonances. The remaining two options are based on experimental signal-to-noise ratio (SNR) as described next.}

\begin{figure}[!htb]
\centering
\includegraphics{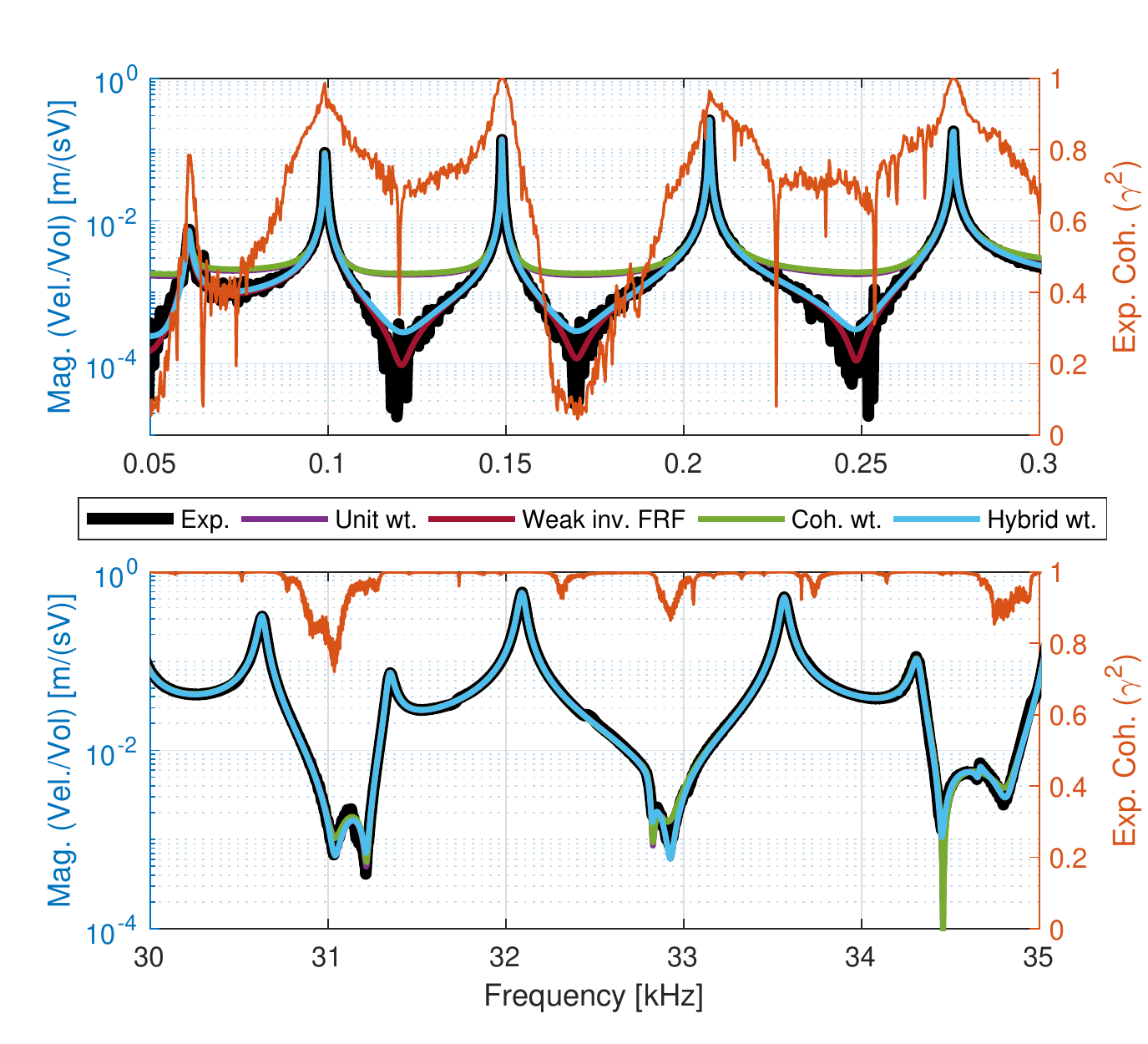}
\caption{ Comparison of the magnitude points of the experimental and the estimated FRFs (at point 6) over two frequency sub-bands of $0.05-0.3~kHz$ and  $30-35~kHz$. The VF estimate of the FRF is generated for the model for all four weighting functions. The magnitude of the experimental FRF is plotted in black. The experimental coherence is plotted in orange. }
\label{fig:comp_weights}
\end{figure}

During the experiments, SNR reduces at anti-resonances.  If the noise floor is higher than the actual FRF value, the measured input and the output signals are incoherent. As a result, the value of the coherence function $\gamma^2(\dot{\imath}\omega)$ drops, indicating low confidence in the measured response. This experimental attribute is dominant at anti-resonances, especially in the lower frequency ranges, as represented by the coherence plots in \Cref{fig:comp_weights}.

\begin{figure}[!htb]
\centering
\includegraphics{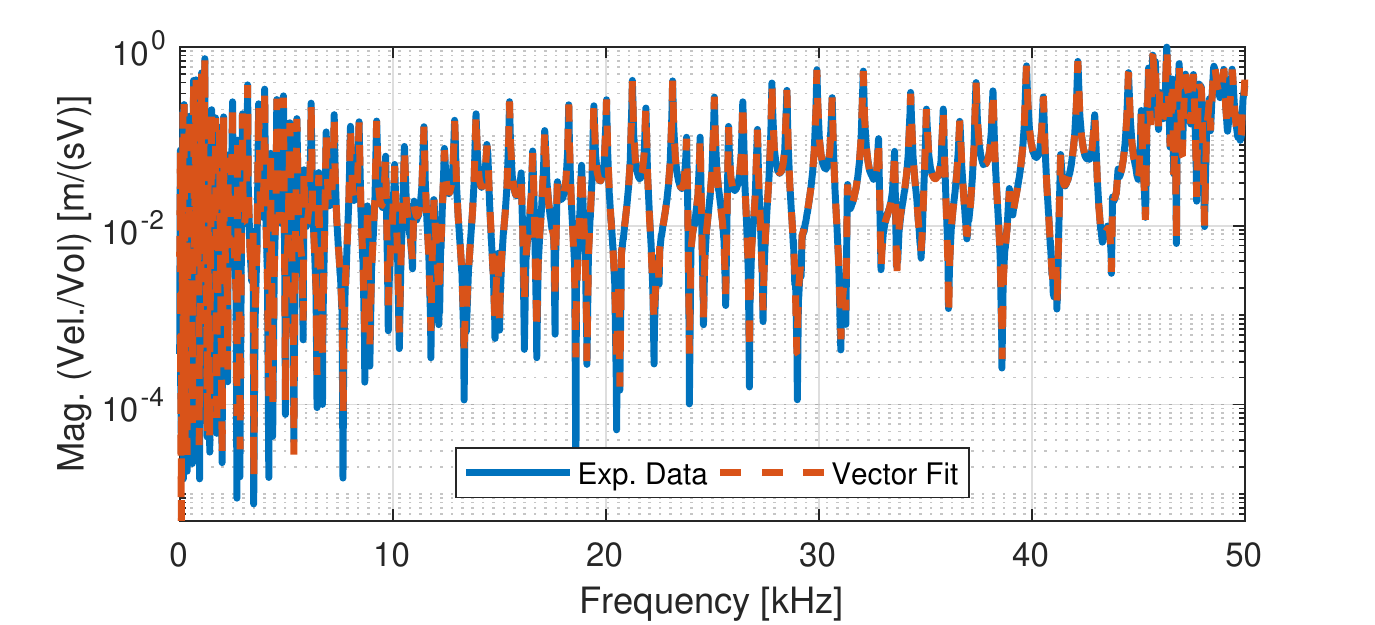}
\caption{ Comparison of the magnitude of the experimental and estimated FRFs (at point 13) over the entire frequency band. The VF estimate of the FRF is generated for the model using hybrid weights. }
\label{fig:Figure_vectorfitFull_Range}
\end{figure}

The remaining two weighting functions make use of this characteristic. Coherence function $\gamma^2(\dot{\imath} \omega_j$) and a hybrid function ($w_j(\dot{\imath} \omega_j)~=~\gamma^2(\dot{\imath} \omega_j)/\sqrt{|H(\dot{\imath} \omega_j)|}$) are the other two weighting functions considered in this study. \Cref{tab:Full_weights} summarizes the performance of the resulting data-driven models, and \Cref{fig:Figure_vectorfitFull_Range} displays the quality of the data-driven model, with the hybrid weighting function, at one of the 19 points (location 13). {The resulting relatives errors  for the four weighting cases are similar.} However, comparing the FRF estimates of all four data-driven models at lower frequencies shows the variation in fitted models. \Cref{fig:comp_weights} compares the out-of-plane FRF estimates at two frequency sub-bands. At higher frequencies, visually, there is little difference between the four FRF estimates.  However, at low-frequency bands, the data-driven models based on the two inverse weighting options perform better in capturing the FRF characteristics at the anti-resonances. While the weak inverse FRF weighting function appears to over-fit noise at anti-resonances, the hybrid weighting function avoids this problem. Section \ref{Sec:dispersion} uses these out-of-plane data-driven models to estimate dispersion curves for the flexural (the first anti-symmetric) wave mode. 

\begin{table}[!htb]
    \centering
    \caption{Performance of VF algorithm in fitting weighted experimental FRFs }
    \begin{tabular}{p{4cm} p{3cm}|p{3cm}p{3cm}}
     \toprule
         \multicolumn{2}{c}{Weighting function $(w_j)$} &  \qquad$\relerr$  & \qquad$\weightrelerr$ \\
         \hline 
         Unit weight &$w_j~=~1$ & ~~~$ 2.63
         \times 10^{-7}$&~~~ $ 2.63\times 10^{-7}$\\[0.7em]
         Weak inv. FRF & $w_j~=~\dfrac{1}{\sqrt{\Hf(\dot{\imath}\omega_j)}}$
         & ~~~$ 3.06\times 10^{-7}$ &~~~$7.48\times 10^{-7}$\\[1.2em]
         Coherence& $w_j~=~\gamma^2(\dot{\imath}\omega_j)$ &~~~$2.64\times 10^{-7}$& ~~~$2.62\times 10^{-7}$\\[1em]
         Hybrid weight &$w_j~=~\dfrac{\gamma^2(\dot{\imath}\omega_j)}{\sqrt{\Hf(\dot{\imath}\omega_j)}}$& ~~~$3.03\times 10^{-7}$&~~~$7.40\times 10^{-7}$\\
         \bottomrule
    \end{tabular}
    \label{tab:Full_weights}
\end{table}


%
\subsection{Vector-fitting In-plane FRFs}

\begin{figure}[!htb]
\centering
\includegraphics{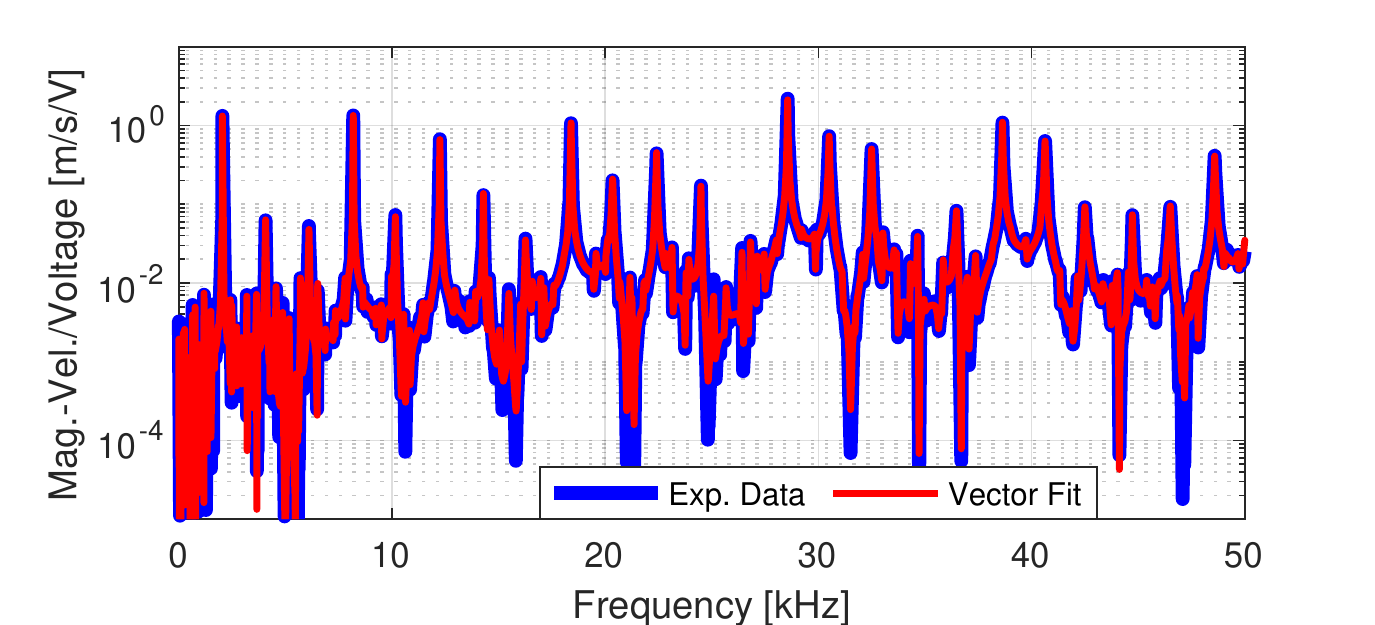}
\caption{Comparison of the magnitude of the experimental and estimated in-plane FRFs (at point 13) over the entire frequency band.}
\label{fig:Figure_vectorfitINPLANEFull_Range}
\end{figure}

This section discusses the development of in-plane data-driven models. Working with in-plane FRFs presents a different set of challenges. In the 1D structure under test, the measured in-plane response is an amalgam of longitudinal modes and flexural/torsional models.  \Cref{fig:Figure_vectorfitINPLANEFull_Range} presents an in-plane experimental FRF. Although longitudinal modes dominate the FRF, the contribution of other vibrations is significant. Limitations in testing are the major contributor to the non-dominant peaks in the experimental FRFs. One way to manage such data is to fit all peaks, including dominant and non-dominant resonant peaks. This approach results in a state-space model of 416 poles. The resultant state-space matrices have the dimensions $\ssAr \in  {\rm I\!R}^{416 \times 416}$, $\ssBr \in  {\rm I\!R}^{416 \times 1}$, and $\ssCr \in  {\rm I\!R}^{19 \times 416}$. Figure \ref{fig:Figure_vectorfitINPLANEFull_Range} presents the experimental FRF along with the fitted FRF. This plot shows that the data-driven FRF accurately captures the dynamics of the longitudinal modes along with artifacts from flexural/torsional vibrations. The error estimate for the $416$ pole data-driven model is  $2.4\times10^{-7}$. 

\section{Estimation of Dispersion Curves using Data-driven Approach}
\label{Sec:dispersion}
\begin{figure}[!htb]
\centering
\includegraphics[trim={0cm 6.9cm 0cm 7cm},clip,width=1\textwidth]{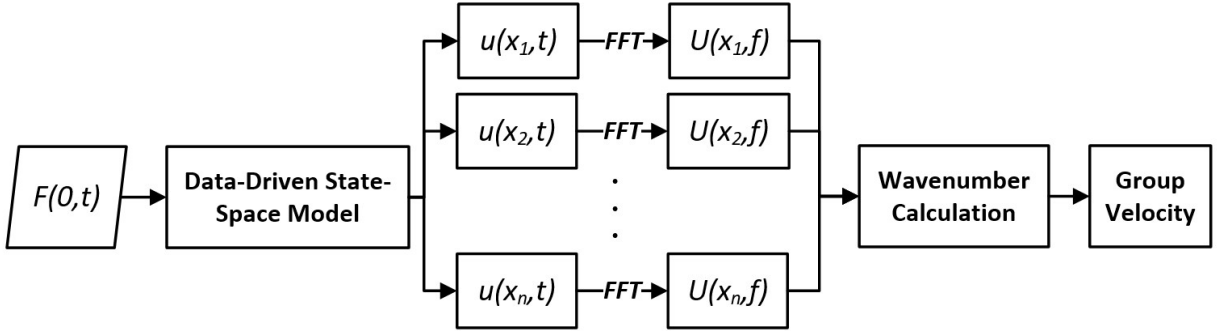}
\caption{A flow chart of the Data-driven-model-based dispersion relations algorithm. In the figure, $F\left( 0,t \right)$ is externally applied external force. $u_i\left( x_i,t \right)$ is simulated response at location $x_i$, and $U_i\left( x_i,f \right)$ is the vector of Fourier coefficients corresponding to $u\left( x_i,t \right)$}.
\label{fig:chart}
\end{figure}

The developed data-driven, state-space model of the beam under test is used to study wave-propagation characteristics along the beam and calculate its dispersion curves. Figure \ref{fig:chart} briefly summarizes the main steps involved in this process. The discussion of the first anti-symmetric (flexural) wave mode is  presented first, followed by the first symmetric (longitudinal) wave mode. A $2$-cycle, amplitude-modulated, sine wave, tone burst is used as the input to the beam's data-driven model. The response to this excitation at all locations where FRFs are experimentally measured is then calculated. Incident waveforms at those locations are extracted from the simulated response, and  the estimated wavenumber, $\hat{k}$, corresponding to the wave mode of interest, is then calculated using the following equation:
\begin{eqnarray}
\label{eqn:Propagation}
\mathbf{U_{i+1}}\left ( \omega \right)=\mathbf{U_i}\left ( \omega \right ) e^{-\dot{\imath}  \mathbf{\hat{k}} \left ( x_{i+1}-x_i \right)},
\end{eqnarray}
where $\mathbf{U_i}$  is the vector of Fourier coefficients of the signal obtained at location $x_i$. Group velocity is then calculated as  $V_G=\partial \omega / \partial \hat{k}$. Dispersion curves are constructed one frequency band at a time by varying the central frequency of the excitation signal and repeating the aforementioned process.

\begin{figure}[!htb]
    \begin{minipage}{0.5\linewidth}\centering
  \includegraphics[width = \linewidth]{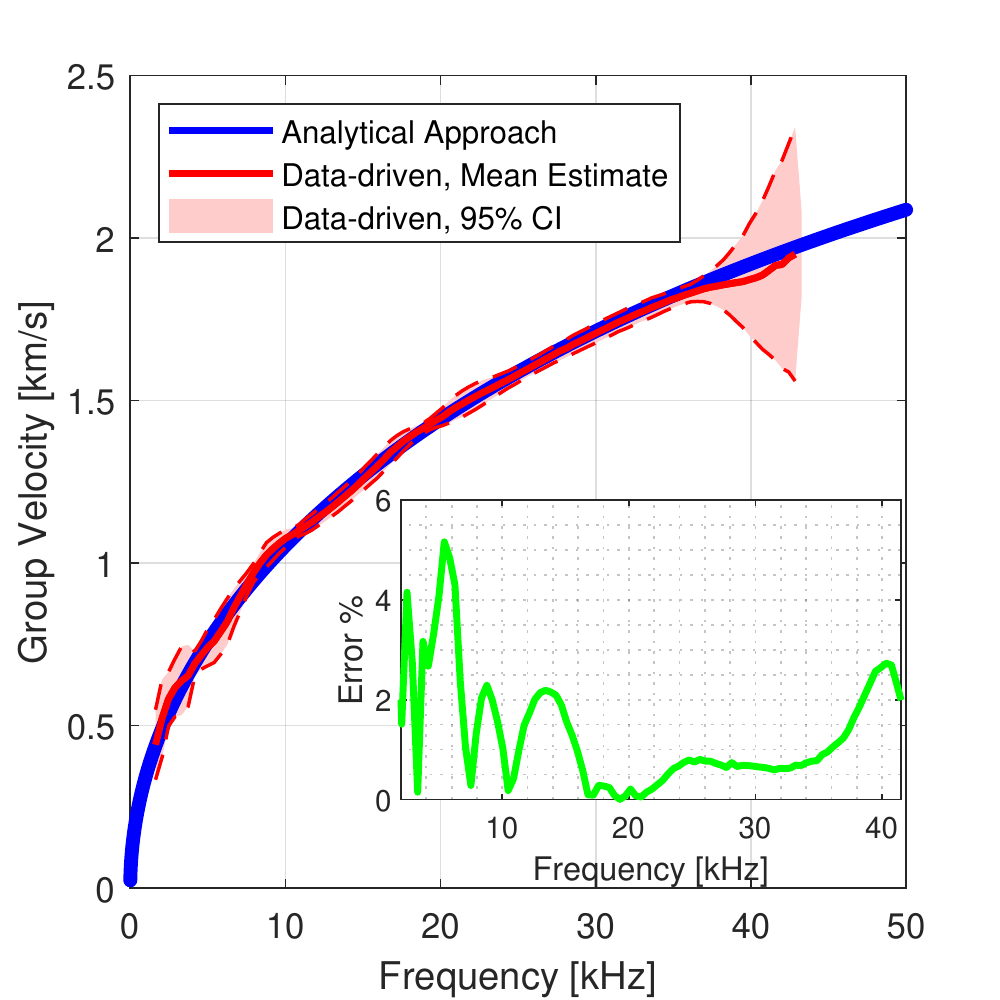}\\
	(a) Unit weight
  \end{minipage}
  \hfill
  \begin{minipage}{0.5\linewidth}\centering
  \includegraphics[width = \linewidth]{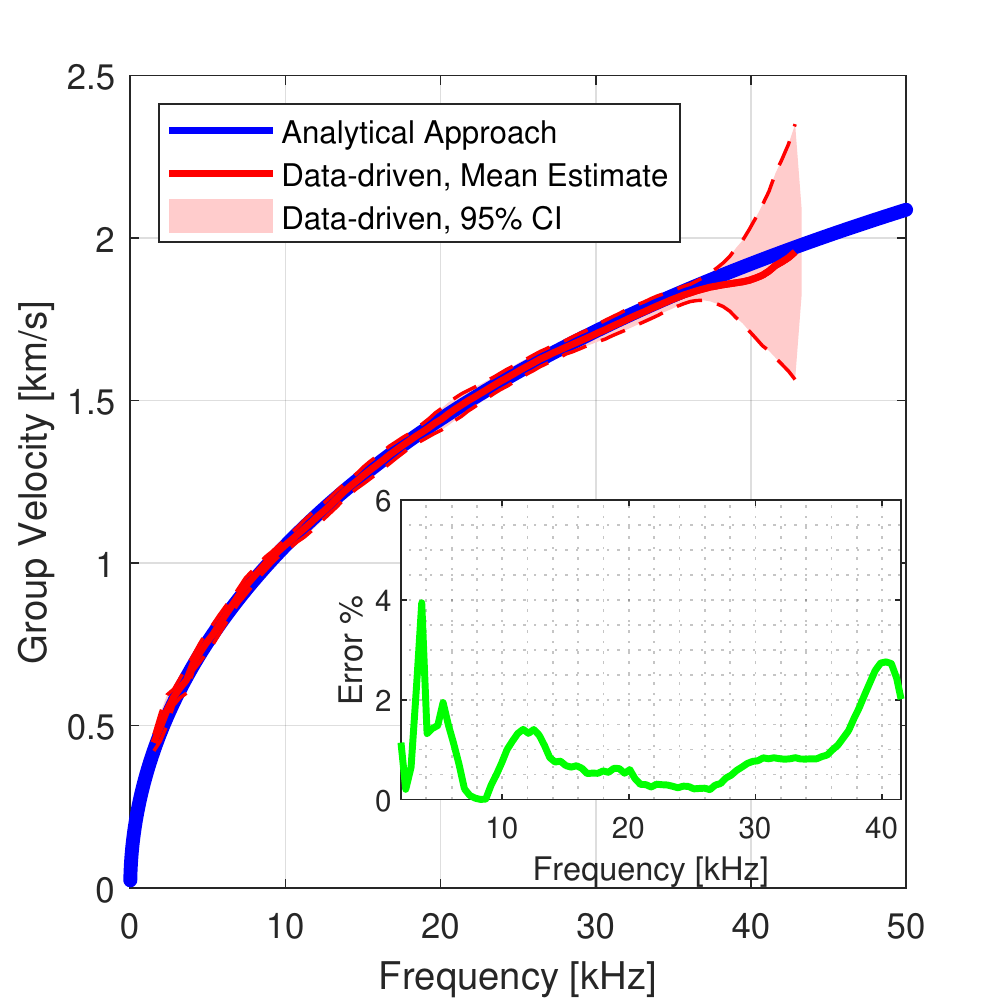}\\
	(b) Weak inverse FRF weight
  \end{minipage}\\
   \begin{minipage}{0.5\linewidth}\centering
  \includegraphics[width = \linewidth]{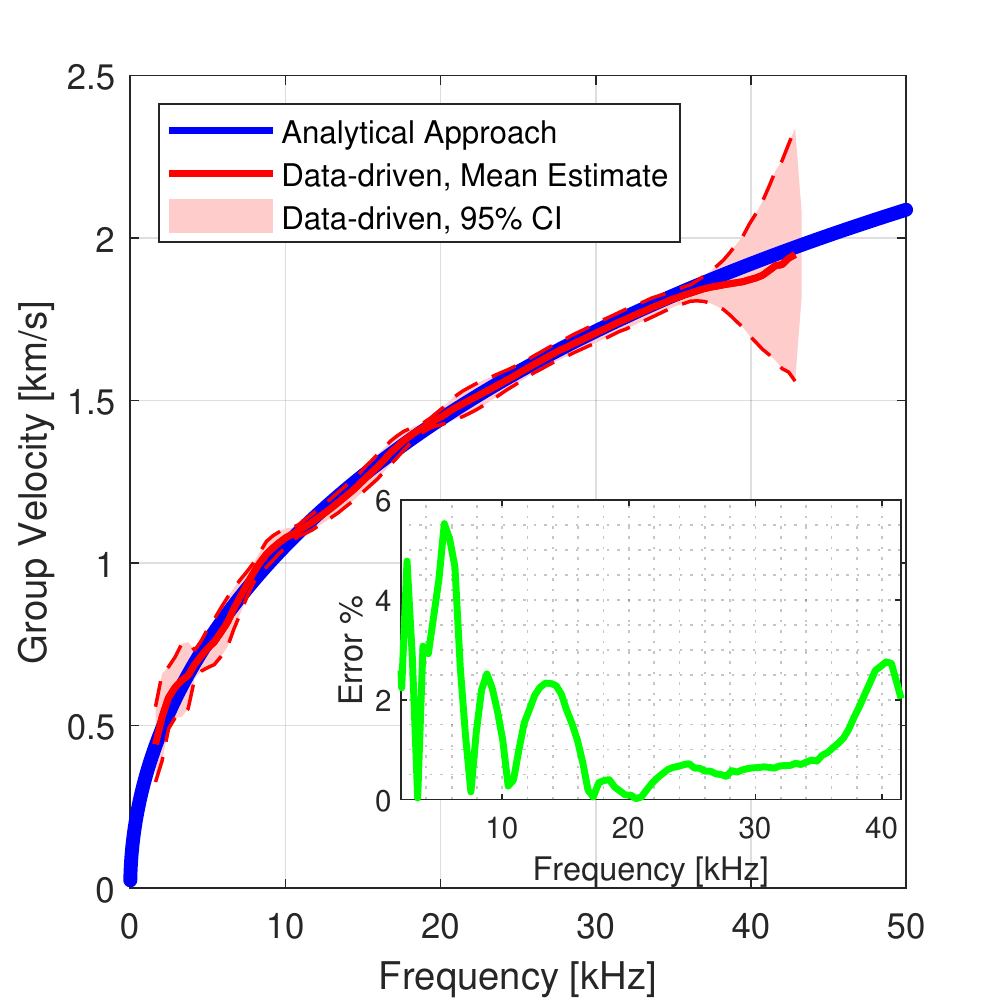}\\
	(c) Coherence weight
  \end{minipage}
  \hfill
  \begin{minipage}{0.5\linewidth}\centering
  \includegraphics[width = \linewidth]{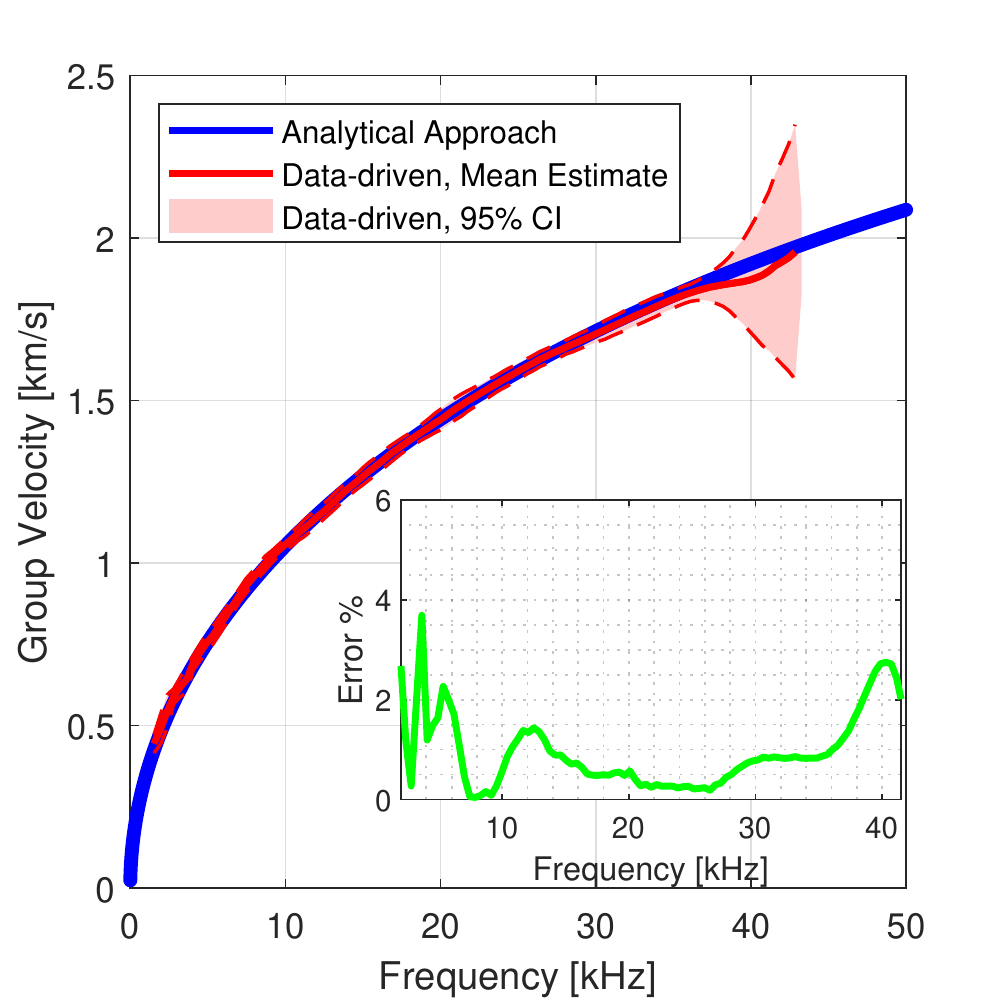}\\
	(d) Hybrid weight
  \end{minipage}
  \caption{ Comparison out-of-plane dispersion curves based on four data-driven models (a) Unit weight, (b) Weak inverse FRF weight, (c) Coherence weight, and (d) Hybrid weight.}
	\label{fig: Flexural_dispersion_curve_Points_weights}
\end{figure}

A minimum of two measurement locations are required to get an estimate of the dispersion curve. However, FRFs at $19$ locations have been obtained and used to build the SIMO  state-space model. This allows for responses to be simulated at those $19$ locations, and thus, a total of $171$ estimates can be obtained. A subset of these combinations is used in this study. Measurement locations closest to the beam's edge are found to be strongly contaminated with reflections at low-excitation frequencies. Thus, these locations are discarded in the analysis. Furthermore, measurement locations are paired such that the distance between them is at least $5-in.$, so that the impact of measurement location uncertainty on dispersion curves estimates is minimized.  \Cref{fig: Flexural_dispersion_curve_Points_weights} shows the results obtained for the flexural wave mode based on the data-driven models developed using the out-of-plane FRFs.

In \Cref{fig: Flexural_dispersion_curve_Points_weights}, the data-driven estimates of the dispersion curves are presented along with the analytical dispersion curve.The analytical curve follows the Timoshenko beam's characteristic equation given by
\begin{align}
    \dfrac{EI}{\rho A} k^4-\dfrac{I}{A}\left(1+\dfrac{E}{G\kappa}\right)k^2\omega^2-\omega^2+ \dfrac{\rho I}{GA\kappa} \omega^4=0,
\end{align}
\noindent where $E~ (70.25\times 10^9 Pa)$ is the Young's modulus of the beam, $\kappa$ is the Timoshenko shear coefficient, $k$ is the theoretical radial wave number, $\rho ~(2650~kg/m^3)$ is the volume density, and $G~(=E/2(1+\nu))$ is the shear modulus and $\nu~(0.33)$ is the Poisson's ratio. The identity $\omega~=k c_p$ gives the analytical dispersion relationship of the beam, in terms of the phase velocity $c_p$. Then, the theoretical group velocity ($c_g$) is given by $c_g=\dfrac{d\omega}{dk}$. \Cref{fig: Flexural_dispersion_curve_Points_weights} compares the experimental (data-driven) estimates of group-velocity ($V_G$) against numerical estimates ($c_g$).

The red-colored curves plotted in \Cref{fig: Flexural_dispersion_curve_Points_weights} are the mean estimate of the group-velocity of flexural waves from data-driven models. The 95\% confidence intervals are also provided for these estimates. The error between the numerical estimates (blue curve) and the data-driven estimates (red curve) is presented via the green curves. The four data-driven models result in four group-velocity curves and four error curves. All four data-driven estimates are close to the analytical curve with a maximum error of $6\%$. In particular, the data-driven models weighted with the weak inverse FRF and the hybrid weight are moderately better at estimating dispersion curves at lower frequencies, with a maximum error under $4\%$. However, all models appear to deviate from the theoretical value in the frequency range over $40~kHz$. This observation in this frequency range agrees with the behavior of the out-of-plane FRFs. Previous notion that the structure's response over $40~kHz$ deviates from a one-dimensional beam is supported by this result. In summary, among all four models, the data-driven model using hybrid weights has the best estimate for the dispersion curve.

\begin{figure}[!htb]\centering
    \begin{minipage}{0.5\linewidth}
  \includegraphics[width = \linewidth]{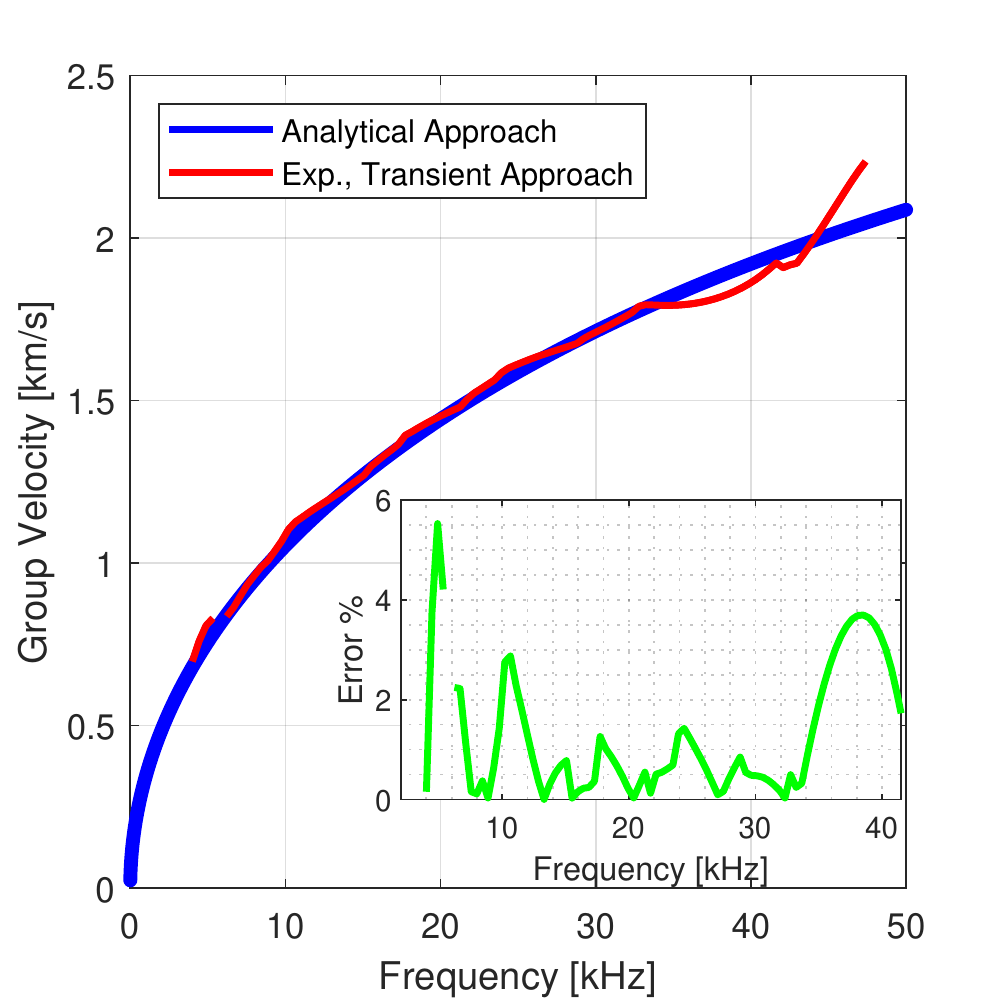}
  \end{minipage}
  \caption{ Comparison between experimentally measured dispersion curves and analytical dispersion curve. The green curve displays the error between the two curves.}
	\label{fig: Flexural_dispersion_curve_Points}
\end{figure}

For comparison purposes, the frequency domain analysis, descried by (\ref{eqn:Propagation}), is applied to the experimentally measured waveforms. These waveforms are obtained with the transient response measurement experiment discussed in Section \ref{Sec:Setup}. Experimentally measured dispersion curves of the beam under test, corresponding to the first anti-symmetric wave mode, are then obtained. The results are depicted in Figure \ref{fig: Flexural_dispersion_curve_Points}. As one can notice, dispersion curves obtained using the proposed data-driven approach (in Figure \ref{fig: Flexural_dispersion_curve_Points_weights}) and the conventional experimental transient-response-based approach (as seen in Figure \ref{fig: Flexural_dispersion_curve_Points}), are in very good agreement. Experimentally measured dispersion curves are also found to agree well with the analytical predictions. With the transient-response-based approach, the high-frequency portion of the dispersion curves has not been measured accurately as reflections from the beam sides strongly contaminated the response. This is the same frequency range where plate-like modes are observed in the steady-state FRFs. Reflections from the beam ends are found to hinder the measurement of the low-frequency portion of the dispersion curve, which is also the case with the proposed data-driven approach. While being an inevitable limitation with the conventional transient-response-based approach, the proposed data-driven approach allow for a number of solutions to be implemented to compensate for reflections.

\begin{figure}[!htb]

  \begin{minipage}[h]{\linewidth}\centering
  \includegraphics[width =0.75 
  \linewidth]{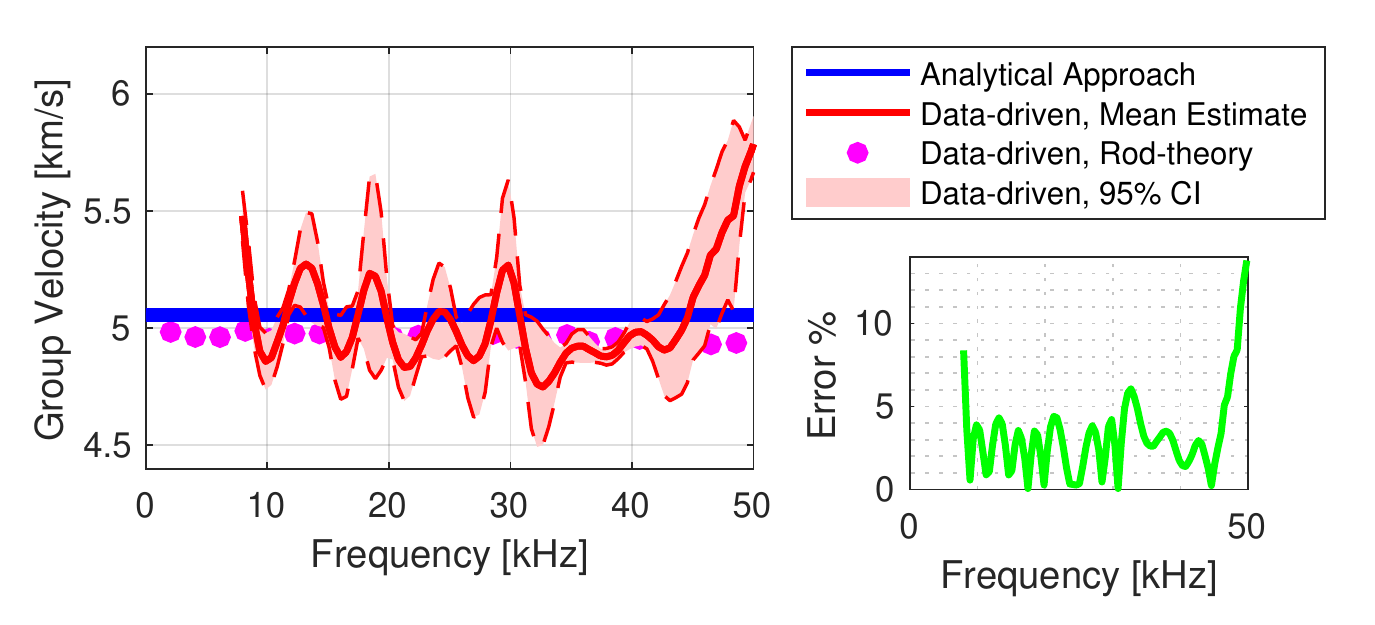}\\
  \end{minipage}
  \caption{Comparison of data-driven dispersion estimates, dispersion estimates based on rod-theory and analytical dispersion curve. The green curve displays the error between the data-driven dispersion estimate and the analytical value.}
	\label{fig: longitudinal_dispersion_curve_Points}
\end{figure}

As for the first symmetric (longitudinal) wave mode, an alternative estimate of the dispersion curve is obtained using the well-known relationship between resonance frequencies and wave speed. For a free-free beam undergoing axial deformations, the elementary rod theory predicts the $n^{th}$ natural frequency of the beam, $\omega_{n}$, to be

\begin{eqnarray}
\label{eqn:Propagation2}
\mathbf{\omega_{n}}=\frac{n\pi c}{L},~~~~~~~~~~n=0,1,2,...
\end{eqnarray}
where $c$ is the wave speed of the first symmetric wave mode and $L$ is the length of the beam under test. The classical rod theory predicts the fist symmetric wave mode to be nondispersive, thus wave speed is frequency independent. Based on the natural frequencies identified over the frequency range of $0-50~kHz$, $24$ estimates of the wave speed can be obtained. The results are depicted in Figure \ref{fig: longitudinal_dispersion_curve_Points}. While a direct wave-speed-estimate based on resonant frequencies provide accurate results, this technique works only when boundary conditions are well-defined. Uncertainty in boundary conditions will reflect on the predicted wave speed. 

\section{Conclusions}
The current work experimentally investigates the performance of a novel data-driven approach in estimating dispersion curves from steady-state FRFs. Experimental data of a free-free beam is adopted as a use-case scenario for this study. A test setup with a Polytec SLDV provides the in-plane and the out-of-plane FRFs for developing data-driven models. Vector fitting the FRFs resulted in two data-driven models:  an out-of-plane model and an in-plane model. These models are qualitatively assessed by comparing the fit at the poles and zeros of the FRFs.  Four weighting functions are considered for improving the quality of the fitted models. Finally, when the FRF data is weighted with a hybrid weighting function, the resulting out-of-plane model with 368 poles {has a relative fitting error in the order of $\mathcal{O}(10^{-7})$. In contrast, the final in-plane model has a similar relative error of order $\mathcal{O}(10^{-7})$ with 416 poles.}

In the next step, the transient response of burst-tone inputs is simulated using the data-driven models. Analyzing the simulated transient response produced the out-of-plane estimates of the phase and group velocities. These data-driven estimates are in good agreement with analytical and experimental (transient) velocities. The out-of-plane data-driven models weighted with the weak inverse FRF and the hybrid weight estimate dispersion curves with a maximum error of under 4\%. Additionally, the data-driven estimates are closer to the analytical values than the transient-experimental estimates. The high-frequency portion of the dispersion curves has not been measured accurately as reflections from the beam sides strongly contaminated the response. This is the same frequency range where plate-like modes are observed in the steady-state FRFs. Reflections from the beam ends are found to hinder the measurement of the low-frequency portion of the dispersion curve, which is also the case with the proposed data-driven approach. While being an inevitable limitation with the conventional transient-response-based approach, the proposed data-driven approach allows for many solutions to be implemented, such as the introduction of artificial damping. This will be further investigated in a future work.

\section*{Acknowledgment}
Tarazaga would like to acknowledge the support provided by the John R. Jones III Faculty Fellowship. The work of Gugercin was supported in parts by NSF through Grant DMS-1819110 and DMS-1720257.


\bibliography{main.bbl}

\end{document}